\documentclass{ifacconf}

\usepackage{graphicx}      
\usepackage{natbib}        
\usepackage{color,amsmath}
\usepackage[latin1]{inputenc}
\usepackage{amsmath}
\usepackage{amssymb} 
\usepackage{tabularx}
\usepackage{color}

\usepackage{subfigure}
\usepackage{epstopdf}
\usepackage{graphicx}

\newtheorem{remark}{Remark}[section]
\begin{document}
\begin{frontmatter}

\title{A simple and efficient feedback control strategy for wastewater denitrification} 


\author[First,Fifth,Sixth]{C\'edric Join} 
\author[Second]{Jean Bernier} 
\author[Third]{St\'ephane Mottelet} 
\author[Fourth,Fifth]{Michel Fliess}
\author[Second]{Sabrina Rechdaoui-Gu\' erin}
\author[Second]{Sam Azimi}
\author[Second]{Vincent Rocher}

\address[First]{CRAN (CNRS, UMR 7039), Universit\'{e} de Lorraine, BP 239, \\ 54506 Vand{\oe}uvre-l\`{e}s-Nancy, France (e-mail: cedric.join@univ-lorraine.fr)}
\address[Second]{SIAAP (Syndicat Interd\'epartemental pour l'Assainissement de l'Agglom\'eration Parisienne), Direction D\'eveloppement Prospective, \\ 82 avenue Kl\'eber, 92700 Colombes, France  \\
(e-mail: \{jean.bernier, sabrina.guerin, sam.azimi, vincent.rocher\}@siaap.fr)}
\address[Third]{TIMR (EA 4297), Sorbonne Universit\'es \& Universit\'e de Technologie de Compi\`egne, rue du docteur Schweitzer, \\ 60203 Compi\`egne, France (e-mail: stephane.mottelet@utc.fr)}
\address[Fourth]{LIX (CNRS, UMR 7161), \'Ecole polytechnique, 91128 Palaiseau, France (e-mail: Michel.Fliess@polytechnique.edu)}
\address[Fifth]{AL.I.E.N. (ALg\`ebre pour Identification \& Estimation Num\'eriques), 24-30 rue Lionnois, BP 60120, 54003 Nancy, France \\ (e-mail: \{cedric.join, michel.fliess\}@alien-sas.com)}
\address[Sixth]{Projet Non-A, INRIA Lille -- Nord-Europe, France}

\begin{abstract}                
Due to severe mathematical modeling and calibration difficulties open-loop feedforward control is mainly employed today for wastewater denitrification, which is a key ecological issue. In order to improve the resulting poor performances a new
model-free control setting and its corresponding ``intelligent'' controller are introduced. The pitfall of regulating two output variables via a single input variable is overcome by introducing also an open-loop knowledge-based control deduced from the plant behavior. Several convincing computer simulations are presented and discussed.

\end{abstract}

\end{frontmatter}
\section{Introduction}
Maintaining low nitrite concentrations in the effluent is a major ecological issue for wastewater treatment plants (WWTP) due to nitrite's high toxicity (see, \textit{e.g.}, \cite{capo,fux,grad,henze,Raimonet2015373,wef}). To this end, the aim of a wastewater post-denitrifying biofilter is to convert nitrate and nitrite ($\mathrm{NO_x,\,x=2,3}$) of the effluent into nitrogen gas (N$_2$). The process uses a submerged packed bed biofilm reactor hosting a class of bacteria under anoxic (low/no oxygen) conditions which use the $\mathrm{NO_x}$ as a source of oxygen when they are fed with a carbon source such as methanol (\cite{Samie2011,Bernier2014}). On the one hand, underfeed of methanol will limit the reduction of $\mathrm{NO_x}$ in the process, and as the denitrification is a two step reaction $\mathrm{NO_3}\rightarrow\mathrm{NO_2}\rightarrow\mathrm{N_2}$, this can leave some nitrite in the effluent even if this compound was absent in the influent (\cite{Rocher2015}). On the other hand, overfeed of methanol results in elevated effluent biochemical oxygen demand (BOD) and useless operating expenses. In most WWTP using post-denitrifying biofilters, the actual control strategy is mainly of feedforward open-loop type:\footnote{See, \textit{e.g.}, \cite{bastin2013line,bourrel2000modelling,crist,predict,ols,tor,wah} for most interesting exceptions, \textit{i.e.}, for the use of feedback loops} on-line measurements of incoming $\mathrm{NO_x}$ are combined with wastewater flow to compute an ideal methanol feed rate. But because of process complexity and many types of disturbances, \textit{e.g.}, periodical backwash of biofilters, this is not sufficient to accurately control the nitrite concentration in the effluent. 
Model-based approaches seem nevertheless to be hard to apply in this context (see, \textit{e.g.}, \cite{grad,wef}). In fact the dynamics of a denitrifying biofilter has to be described by partial differential equations taking into account concentration gradients, nonlinearities of the biological processes, biofilm clogging and so on (see, \textit{e.g.}, the excellent report by \cite{pde}, and the references therein). Such an exhaustive model needing the identification of numerous parameters is necessary to assess a particular control strategy, like in Section \ref{sec:SC}. However it cannot reasonably be used for realtime feedback control. 

A new \emph{model-free control} setting (\cite{ijc}) is therefore used.\footnote{See, \textit{e.g.}, \cite{edf} for the analysis of model-free control in a somehow analogous situation with respect to partial differential equations.} The corresponding \emph{intelligent} P feedback controller is moreover quite easy to implement both from software (\cite{ijc}) and hardware (\cite{nice}) standpoints. Many concrete applications have already been developed all over the world. Some have been patented. Lack of space permits here to quote only three recent works that are related to biotechnology: \cite{bara,toulon,med16}. From a purely control-theoretic viewpoint, a major difficulty is encountered: a single input variable must regulate two output variables (see, \textit{e.g.}, \cite{diff} for an explanation via input-output nonlinear system inversion). A satisfactory practical solution is nevertheless proposed.  It is based on an appropriate mixing of a knowledge-based behavior of the plant with a suitable feedback law. 

Our paper is organized as follows. Section \ref{model} is sketching wastewater treatment and its corresponding rough description via differential equations. Model-free control and the associated real-time estimation techniques are summarized in Section \ref{mfc}. After presenting our control law, Section \ref{numer} displays convincing computer experiments, which are based in this preliminary work on the modeling in Section \ref{model}. Some concluding remarks may be found in Section \ref{con}.

\section{Problem and model}\label{model}
\begin{figure}[t]
\includegraphics[width= \linewidth]{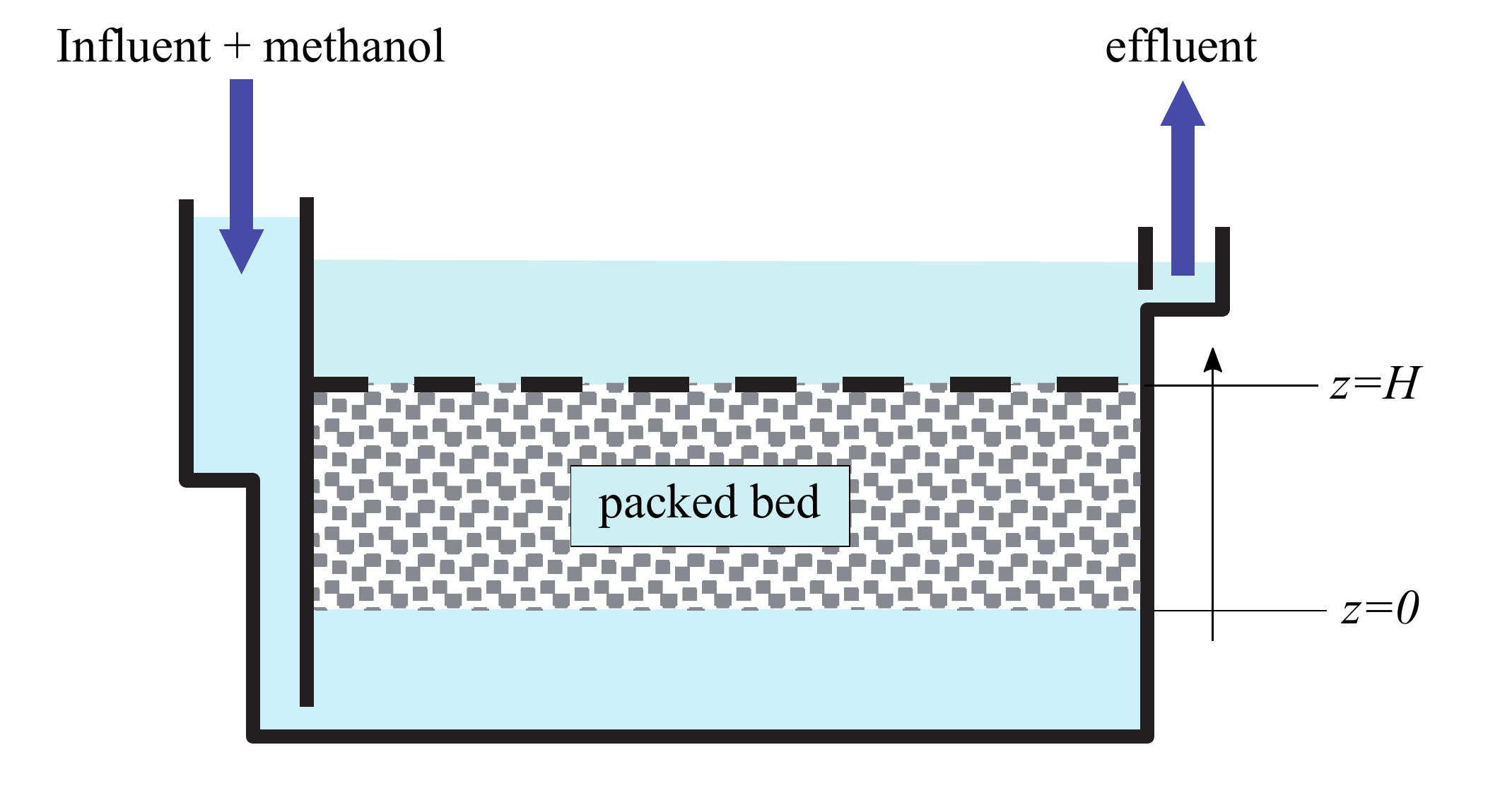}
\caption{Packed bed denitrifying biofilter}
\label{fig:biofilter}
\end{figure}
\subsection{Dynamic model of the biofilter}\label{dyn}
The actual denitrifying process is composed of several biofilters operating in parallel with the same influent. A typical unit is displayed in Figure \ref{fig:biofilter}. The water is fed from the base of the filter bed, which is composed of beads of expanded clay. A biofilm, where all biological reactions occur, is considered to grow on the media. The biological reactions inside the biofilm are modeled using a modified version of ASM1 (\cite{henzeASM1}). This model is widely used to simulate the growth of bacteria and the resulting consumption of carbon and nitrogen pollution occurring in biological wastewater treatment processes. The main modification in this specific case consists in the addition of two-step denitrification to simulate the production and consumption of nitrite during the process. The first two denitrification reactions of ASMN (\cite{hiatt}) were used to this end, and can be summarized as 
\begin{align*}
\mathrm{methanol}+\mathrm{nitrates} &\rightarrow \mathrm{biomass}+\mathrm{nitrites},\\
\mathrm{methanol}+\mathrm{nitrites} &\rightarrow \mathrm{biomass}+\mathrm{nitrogen}.
\end{align*}
The whole dynamics of the biofilter results from the mass balance of the different reacting species, nitrate $S_1$, nitrite $S_2$, carbon $S_C$ and biomass $X$. The limited axial dispersion allows to consider that all concentrations are constant in a vertical cross section. The following model with distributed parameters, \textit{i.e.}, a system of partial differential equations, model with a single space dimension, may be written
\begin{align}
\label{eq:model1}\varepsilon\partial_t S_1 &=-v\partial_zS_1-k_1\mu_1(S_1,S_c)X\\
\varepsilon\partial_t S_2 &=-v\partial_zS_2+k_1\mu_1(S_1,S_c)X-k_2\mu_2(S_2,S_c)X\\
\label{eq:model3}\varepsilon\partial_t S_c &=-v\partial_zS_c-k_3\mu_1(S_1,S_c)X-k_4\mu_2(S_2,S_c)X\\
\label{eq:model4} \partial_t X&=(\mu_1(S_1,S_c)+\mu_2(S_2,S_c))(1-X/X_{\text{max}})X
\end{align}
for $z\in]0,H]$. All species concentrations are given in g/m$^3$. The yield coefficients $(k_i)_{i=1, \dots,4}$ are given in $g/g$, the superficial velocity $v$ in m/h (flow rate in m$^3$/h divided by the cross-section area) and $\varepsilon$ is the porosity. The specific growth rates $\mu_1$ and $\mu_2$ are given by a double Monod-type model (\cite{bastin2013line})
$$\mu_i(S_i,S_c)=\mu_{i,\text{max}}\frac{S_iS_c}{(K_i+S_i)(K_c+S_c)}\quad i=1,2$$
where $\mu_{i,\text{max}}$ is the maximum specific growth rate for species $S_i$ and $K_i,K_c$ the affinity constants. The boundary conditions are given by
$$
S_1(0,t)=S_{1,in}(t), \,S_2(0,t)=S_{2,in}(t),\,S_c(0,t)=S_{c,in}(t)
$$
where $S_{c,in}(t)$ is the methanol concentration at the inlet of the reactor, i.e. the control variable.

Equations (\ref{eq:model1})-(\ref{eq:model4}) were initially proposed for a drinkable water denitrifying biofilter (see (\cite{bourrel2000modelling}) and were easily adapted to our specific configuration as only hydraulic parameters and maximum biomass concentration $X_{\text{max}}$ had to be changed. They have been used during the early stage of our project in order to validate our interest for model-free control, but the simulations of Section \ref{sec:SC} have been made with the \emph{SimBio} software. This more realistic setting is built in Matlab with the Simulink toolbox, using submodels already available in the literature. The biofilter hydraulics is approximated with a series of several continuously stirred tank reactors (CSTRs) of equal volume to obtain reactor hydraulics close to the plug-flow model of equations (\ref{eq:model1})-(\ref{eq:model3}) while maintaining simulation times in a reasonable range. As concentration gradients are observed in thick biofilms, their distributed nature is taken into account in \emph{SimBio} by dividing the biofilm into several CSTRs through which soluble substrates are able to diffuse (\cite{spengel}). Soluble substrates are brought to and into the biofilm through diffusion, whereas particular components are transferred to the biofilm surface through filtration (\cite{horner,ives}). Backwash efficiency is modelled as a removal of a fixed proportion of biofilm thickness in each reactor using different removal efficiencies for biomass and for other non-biomass particles. A certain fraction of media mixing across the reactors also occurs during backwash. 

This version of \emph{SimBio} was calibrated on hourly nitrate and nitrite measurements made on the post-denitrification step of the Seine-Centre plant (\cite{Bernier2014}).%
\subsection{Actual control strategy on the real plant}
The control variable value $S_{c,in}(t)$ is tuned according to a desired removal of incoming nitrogen, which comes under the form of nitrates (the incoming concentration of nitrite is negligible), \textit{i.e.}, for a target concentration of $S_{1,\text{target}}$ at the outlet of the reactor, the control law is of the form
\begin{equation}
S_{c,in}(t)=\beta(S_{1,in}(t)-S_{1,\text{target}})
\label{eq:fflaw}
\end{equation}
where $\beta$ is an operating coefficient based on pure stoichiometric and yield considerations (\cite{Rocher2015}). This strategy does not take into account the intermediate species (nitrite) and leads to an unstable behavior of its concentration in the effluent. In fact the ratio between the methanol and the total nitrogen (nitrate and nitrite) concentrations in the biofilter plays a major role in the appearance of residual nitrites, but it cannot be controlled with a simple control law as in Equation (\ref{eq:fflaw}). However, as the nitrite concentration can be measured at the outlet of the biofilter, it can be used as a natural controlled variable in a feedback control strategy.

\section{Model-free control}\label{mfc}
\subsection{The ultra-local model}
Replace the unknown global system description by the \emph{ultra-local model}:\footnote{For more details, see \cite{ijc}.}
\begin{equation}
\dot{y} = F + \alpha u \label{1}
\end{equation}
where
\begin{itemize}
\item the control and output variables are $u$ and $y$,
\item the derivation order of $y$ is $1$ like in most concrete situations,
\item $\alpha \in \mathbb{R}$ is chosen by the practitioner such that $\alpha u$ and
$\dot{y}$ are of the same magnitude.
\end{itemize}
The following explanations on $F$ might be useful: 
\begin{itemize}
\item $F$ is estimated via the measures of $u$ and $y$,
\item $F$ subsumes not only the unknown system structure but also
any perturbation.
\end{itemize}
\begin{remark}
In Equation \eqref{1} $\dot y$ is seldom replaced by $\ddot y$ (see, \textit{e.g.}, \cite{ijc}, and the references therein). Higher order derivatives were never utilized until today.
\end{remark}
\subsection{Intelligent controllers}
The loop is closed by an \emph{intelligent proportional controller}, or \emph{iP},
\begin{equation}\label{ip}
u = - \frac{F - \dot{y}^\ast + K_P e}{\alpha}
\end{equation}
where
\begin{itemize}
\item $y^\star$ is a reference trajectory,
\item $e = y - y^\star$ is the tracking error,
\item $K_P$ is the usual tuning gain.
\end{itemize}
Combining Equations \eqref{1} and \eqref{ip} yields:
$$
\dot{e} + K_P e = 0
$$
where $F$ does not appear anymore. The tuning of $K_P$, in order to insure local stability, becomes therefore quite straightforward. This is a major benefit when
compared to the tuning of ``classic'' PIDs (see, \textit{e.g.},
\cite{astrom,murray}, and the references therein), which
\begin{itemize}
\item necessitate a ``fine'' tuning in order to deal with the poorly known parts of the plant,
\item exhibit a poor robustness with respect to ``strong'' perturbations and/or system alterations.
\end{itemize}

\subsection{Estimation of $F$}\label{F}
The calculations below stem from algebraic estimation techniques that are borrowed from \cite{sira1,sira2}, and \cite{sira}.

\subsubsection{First approach}
The term $F$ in Equation \eqref{1} may be assumed to be ``well'' approximated by a piecewise constant function $F_{\text{est}} $ (see, \textit{e.g.}, \cite{godement}). Rewrite then Equation \eqref{1}  in the operational domain (see, \textit{e.g.}, \cite{yosida}): 
$$
sY = \frac{\Phi}{s}+\alpha U +y(0)
$$
where $\Phi$ is a constant. We get rid of the initial condition $y(0)$ by multiplying both sides on the left by $\frac{d}{ds}$:
$$
Y + s\frac{dY}{ds}=-\frac{\Phi}{s^2}+\alpha \frac{dU}{ds}
$$
Noise attenuation is achieved by multiplying both sides on the left by $s^{-2}$. It yields in the time domain the realtime estimate, thanks to the equivalence between $\frac{d}{ds}$ and the multiplication by $-t$,
\begin{equation}\label{integral1}
{\small F_{\text{est}}(t)  =-\frac{6}{\tau^3}\int_{t-\tau}^t \left\lbrack (\tau -2\sigma)y(\sigma)+\alpha\sigma(\tau -\sigma)u(\sigma) \right\rbrack d\sigma} 
\end{equation}

\subsubsection{Second approach}\label{2e}
Close the loop with the iP \eqref{ip}:
\begin{equation}\label{integral2}
F_{\text{est}}(t) = \frac{1}{\tau}\left[\int_{t - \tau}^{t}\left(\dot{y}^{\star}-\alpha u
- K_P e \right) d\sigma \right] 
\end{equation}
\begin{remark}
Note the following facts: 
\begin{itemize}
\item integrals \eqref{integral1} and \eqref{integral2} are low pass filters,
\item $\tau > 0$ might be quite small,
\item the integrals may of course be replaced in practice by classic digital filters.
\end{itemize}
\end{remark}

\section{Numerical experiments} \label{numer}
\label{sec:SC}
\subsection{Control law description}\label{des}
\begin{figure}[t]
\center\includegraphics[width=\linewidth]{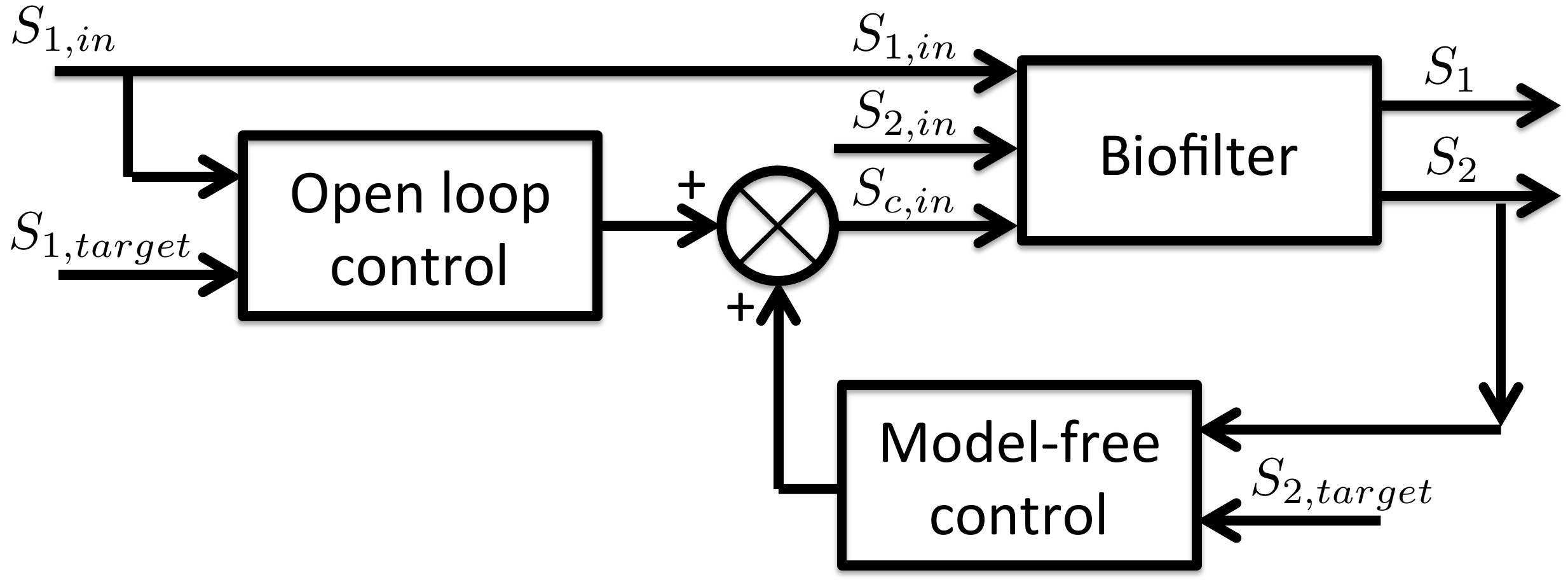}
\caption{Control scheme}
\label{fig:scheme}
\end{figure}

Our control law is displayed by Figure \ref{fig:scheme}:
\begin{itemize}
\item Equation \eqref{eq:fflaw} defines the single control variable in open loop,
\item the loop is closed via the iP \eqref{ip}.
\end{itemize}
The aim of methanol injection $S_{c,\text{in}}$, \textit{i.e.}, our single input variable,\footnote{See Section \ref{dyn} for precise definitions of the quantities associated to the letter $S$.} is to regulate the total nitrogen concentration, \textit{i.e.}, two quantities $S_{1}$ and $S_{2}$, namely the nitrate and nitrite wastewater concentrations. The fact, depicted in Figure 3, that $S_1$ is much larger than $S_2$, is taken into account by regulating $S_1$ via an open-loop knowledge-based control. For $S_2$ model-free control is utilized. For Equations \eqref{1}-\eqref{ip}, $\alpha=1$ and $K_p=100$ were selected. In order to avoid debating the regulation of $S_1$ the resulting value of the iP should be non-negative. If not, the water concentration of nitrate would increase. This is not acceptable. 


\subsection{Simulations} 
The mathematical modeling discussed in Section \ref{dyn} is used for the computer simulations. The sampling time period is $0.001$ day.   

Quite good results corresponding to several targets are displayed in Figures \ref{S04}, 5, 6, 7, \ref{S30}. Important daily perturbations, corresponding to biofilter backwash, have been introduced in order to show more realistic performances. According to the Figures, a small injection of methanol is reducing notably the nitrites and nitrates concentrations in the water which is rejected. Notice however a worsening of the performances 
if $S_{2,\text{target}}\geq 1.2$. Let us emphasize that 
\begin{itemize}
\item it is not due to a weakness of our control strategy,
 \item the very nature of the wastewater denitrification, which is detailed at the end of Section \ref{des}, explains it.
\end{itemize}



%

\begin{figure*}
\begin{center}
\subfigure[Nitrate $S_{1,in}$]{
\resizebox*{8cm}{!}{\includegraphics{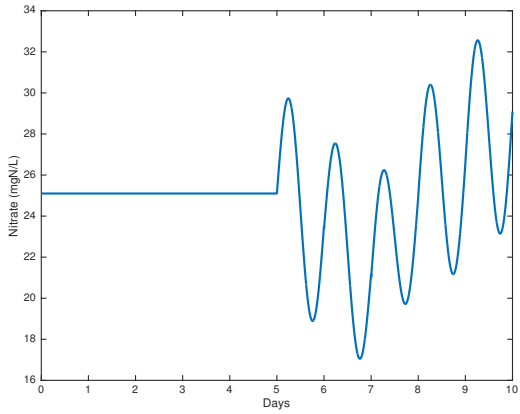}}}%
\subfigure[Nitrite $S_{2,in}$]{
\resizebox*{8cm}{!}{\includegraphics{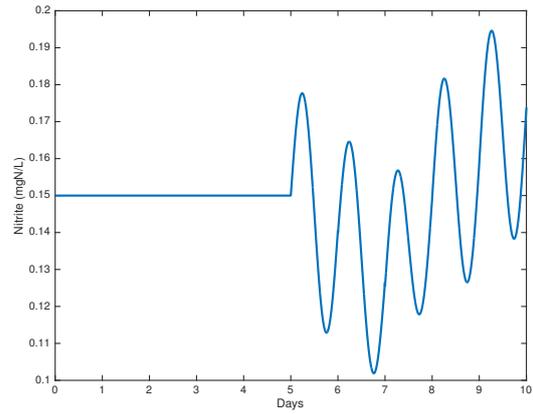}}}%
\caption{Influent}
\label{Sin}
\end{center}
\end{figure*}

\begin{figure*}
\begin{center}
\subfigure[Nitrate $S_1$]{
\resizebox*{5.22cm}{!}{\includegraphics{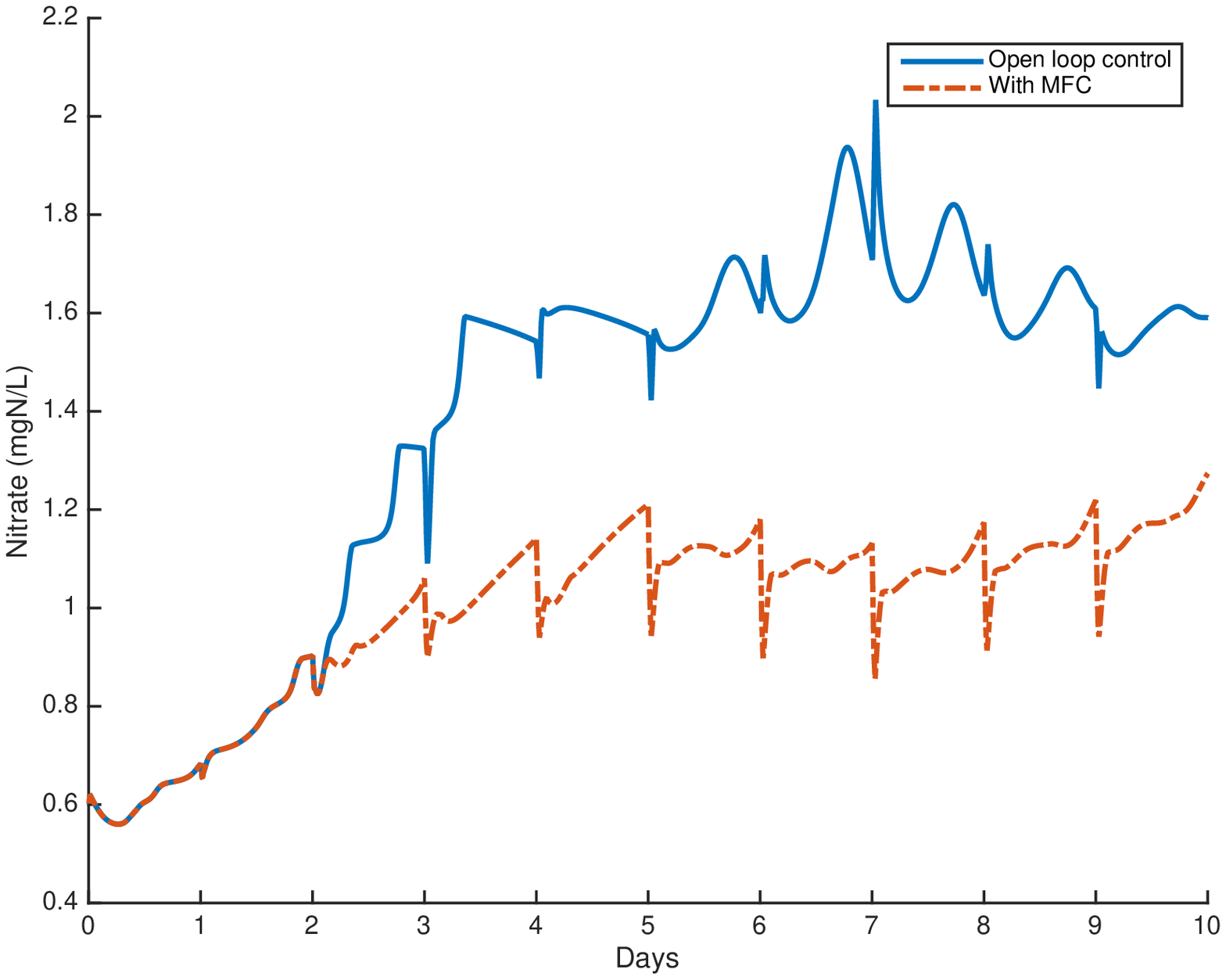}}}%
\subfigure[Nitrite $S_2$]{
\resizebox*{5.22cm}{!}{\includegraphics{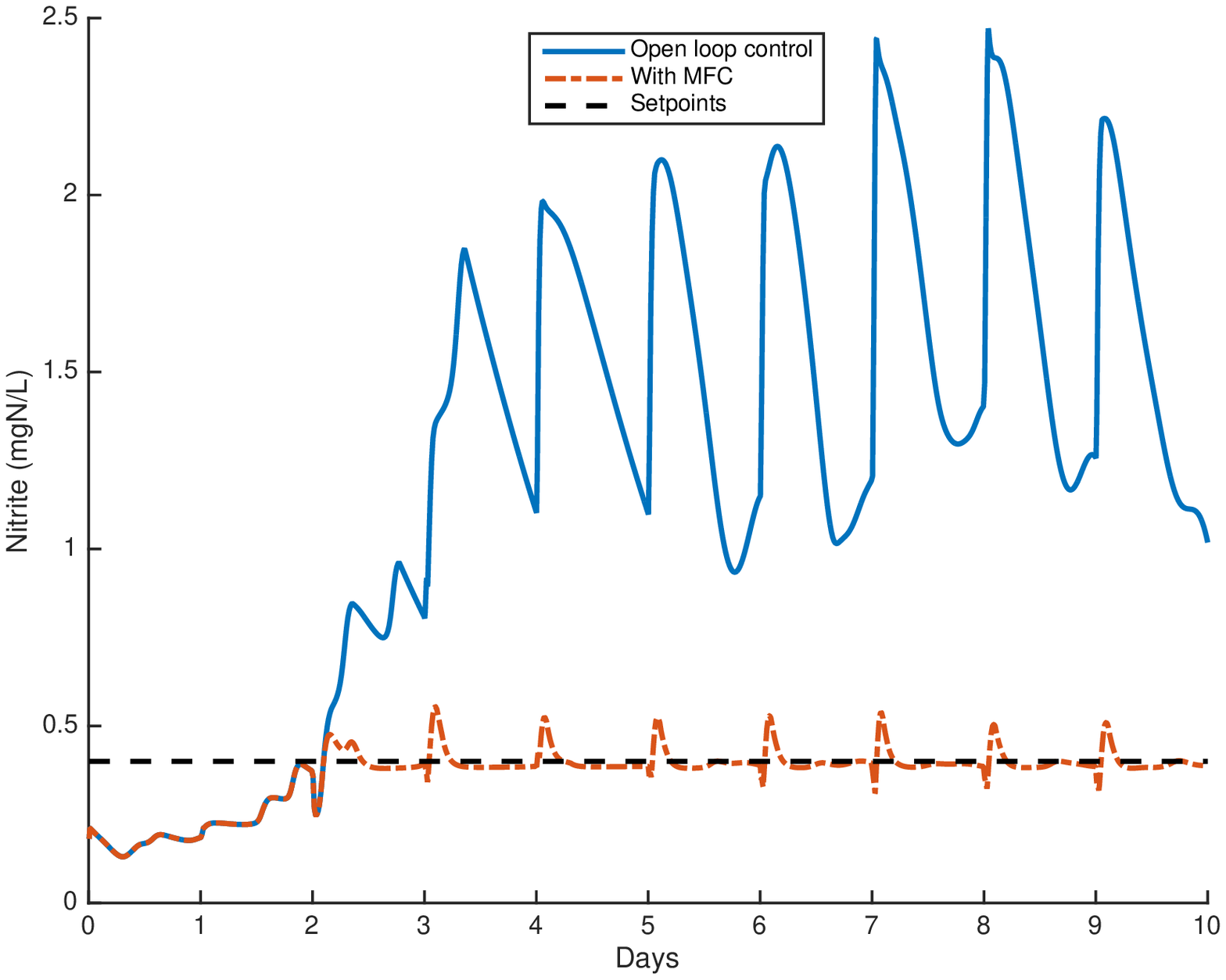}}}%
\subfigure[Methanol $S_{c,in}$]{
\resizebox*{5.22cm}{!}{\includegraphics{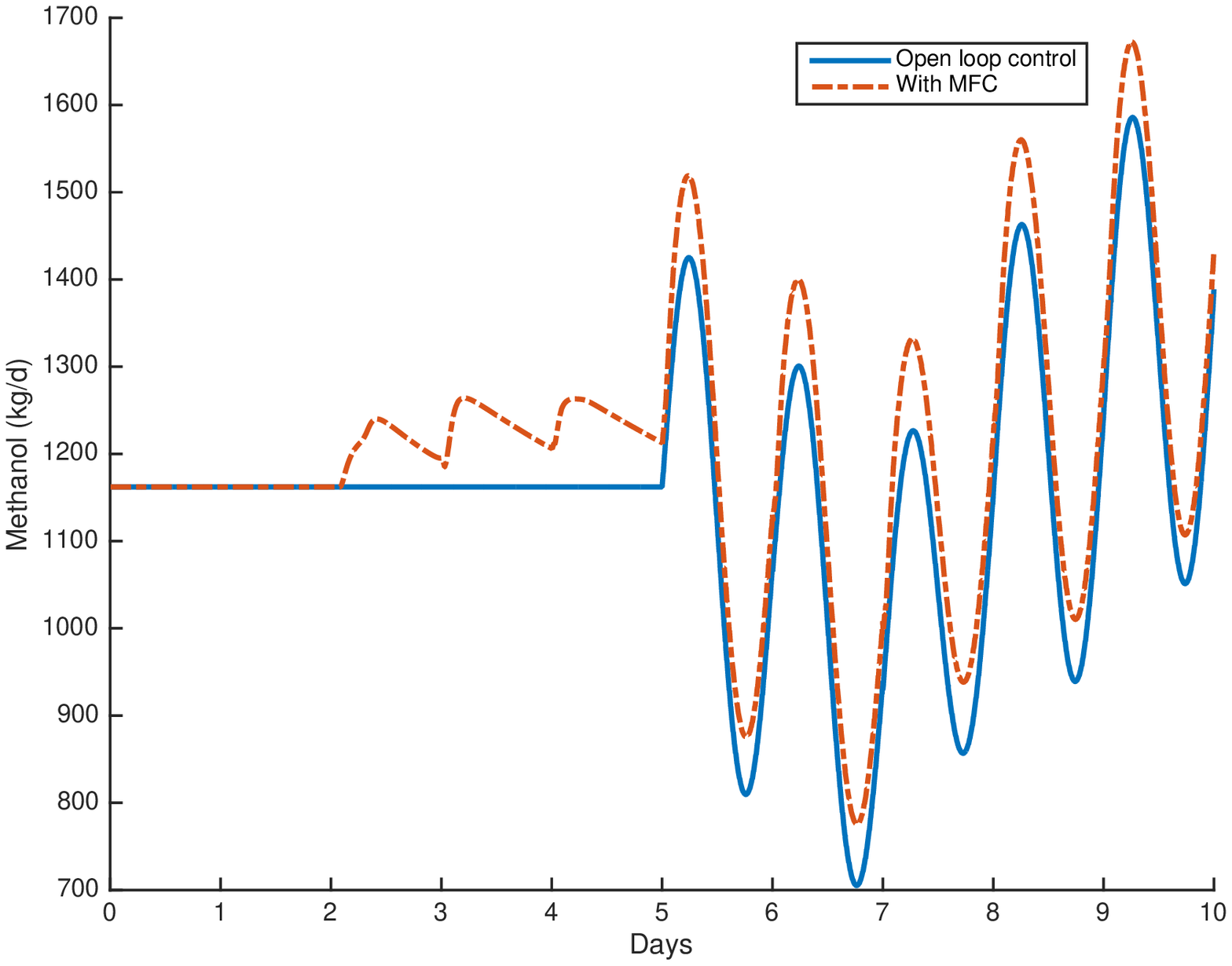}}}%
\caption{Setpoint : $S_{2,\text{target}}=0.4$ }%
\label{S04}
\end{center}
\end{figure*}

\begin{figure*}
\begin{center}
\subfigure[Nitrate $S_1$]{
\resizebox*{5.22cm}{!}{\includegraphics{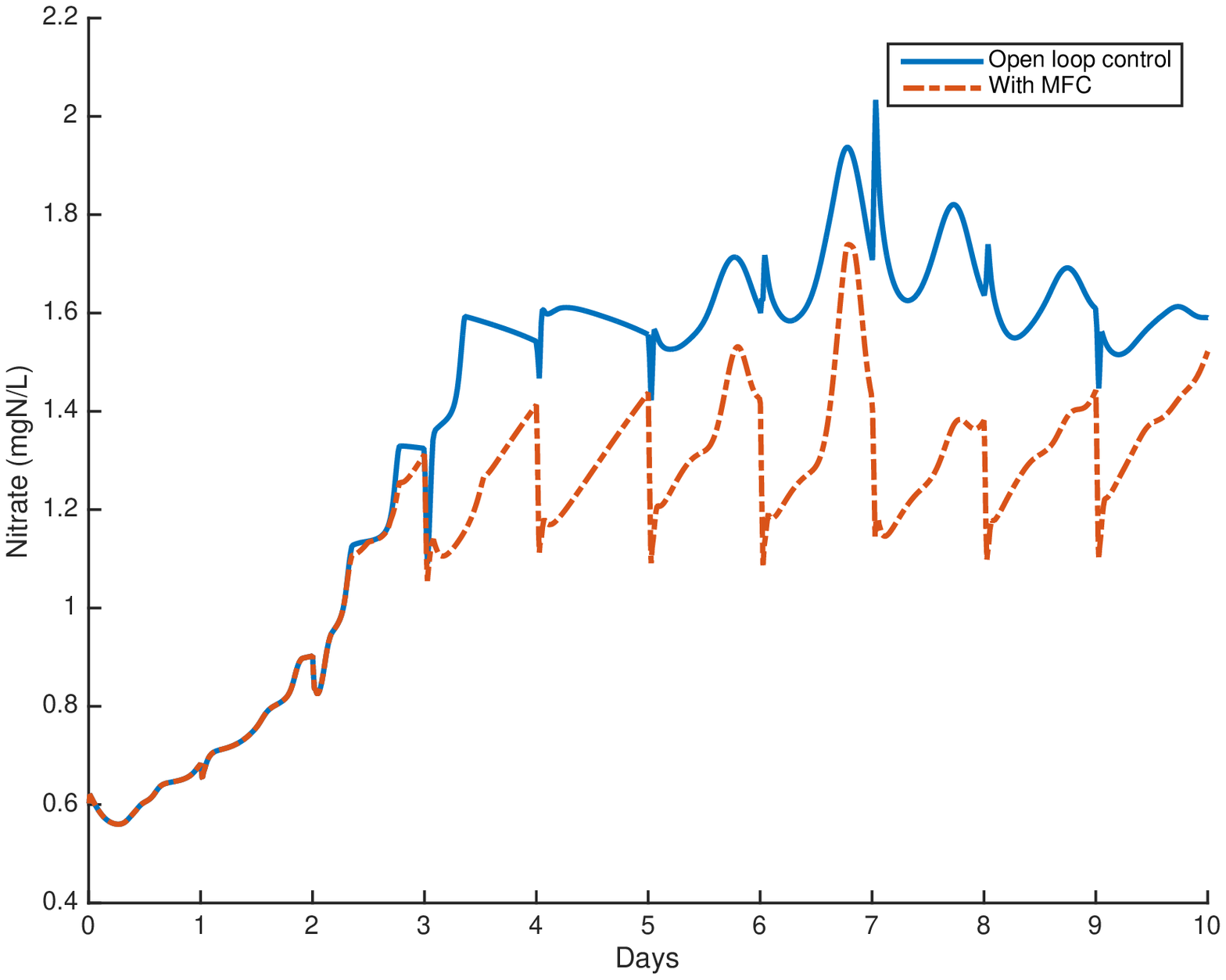}}}%
\subfigure[Nitrite $S_2$]{
\resizebox*{5.22cm}{!}{\includegraphics{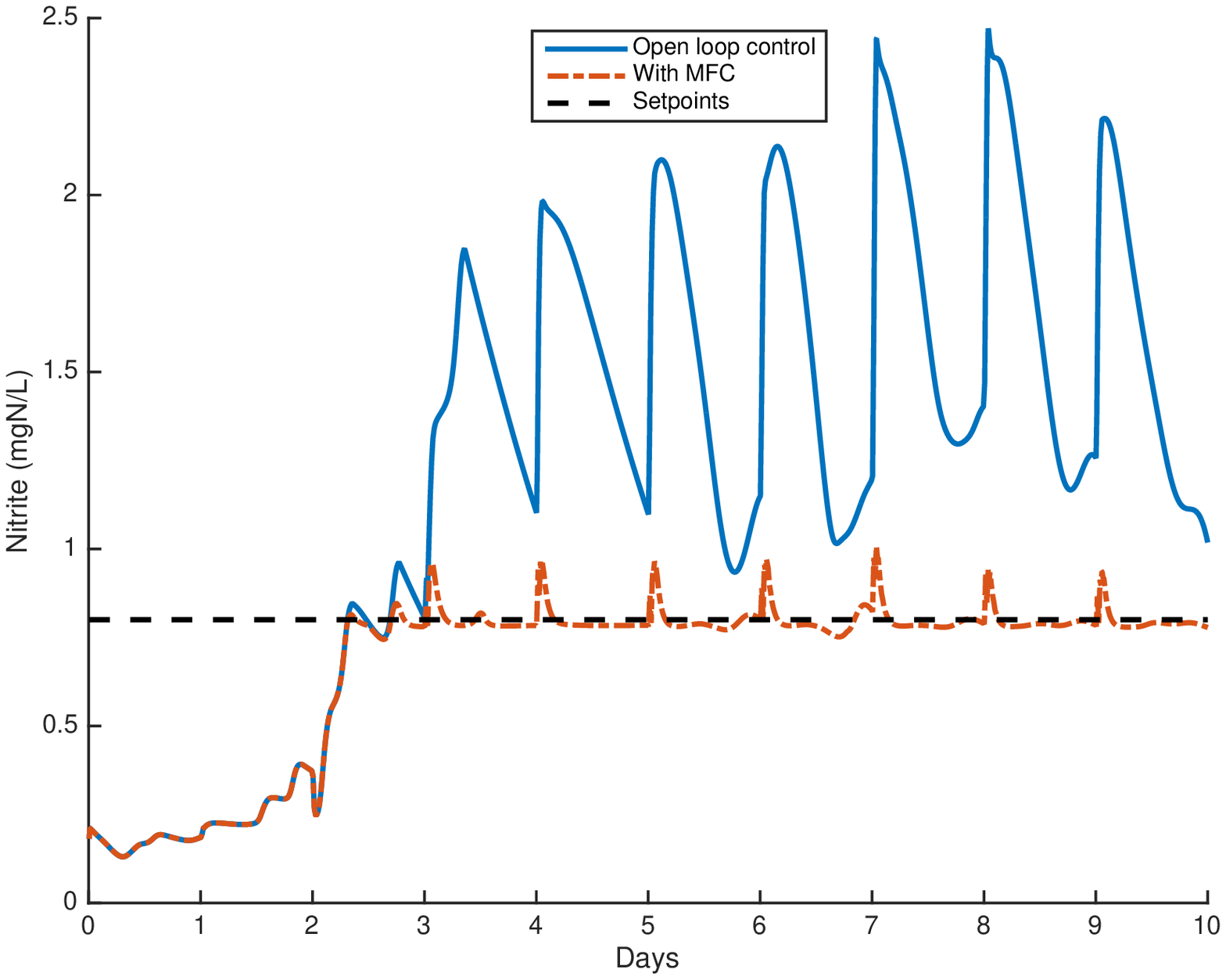}}}%
\subfigure[Methanol $S_{c,in}$]{
\resizebox*{5.22cm}{!}{\includegraphics{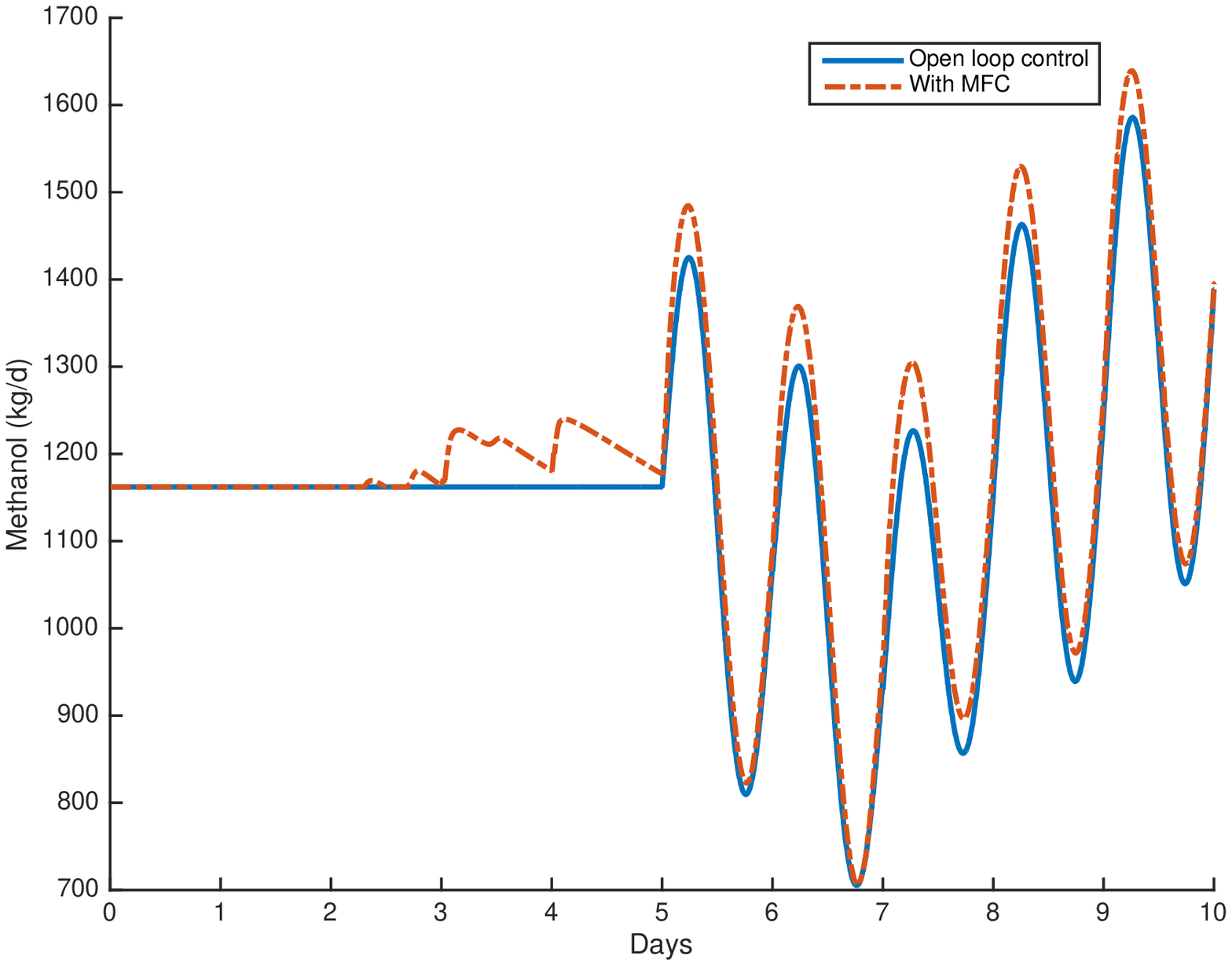}}}%
\caption{Setpoint : $S_{2,\text{target}}=0.8$ }%
\label{S08}
\end{center}
\end{figure*}

\begin{figure*}
\begin{center}
\subfigure[Nitrate $S_1$]{
\resizebox*{5.22cm}{!}{\includegraphics{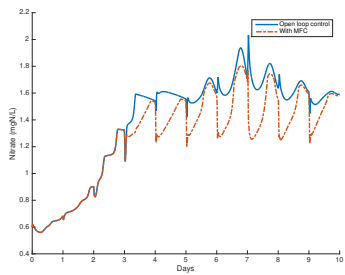}}}%
\subfigure[Nitrite $S_2$]{
\resizebox*{5.22cm}{!}{\includegraphics{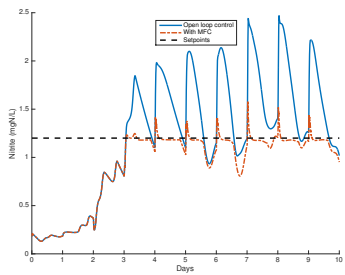}}}%
\subfigure[Methanol $S_{c,in}$]{
\resizebox*{5.22cm}{!}{\includegraphics{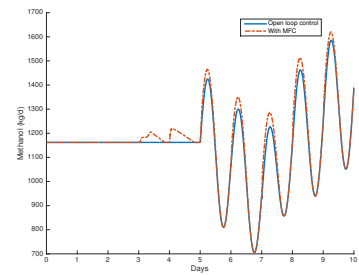}}}%
\caption{Setpoint : $S_{2,\text{target}}=1.2$ }%
\label{S12}
\end{center}
\end{figure*}

\begin{figure*}
\begin{center}
\subfigure[Nitrate $S_1$]{
\resizebox*{5.22cm}{!}{\includegraphics{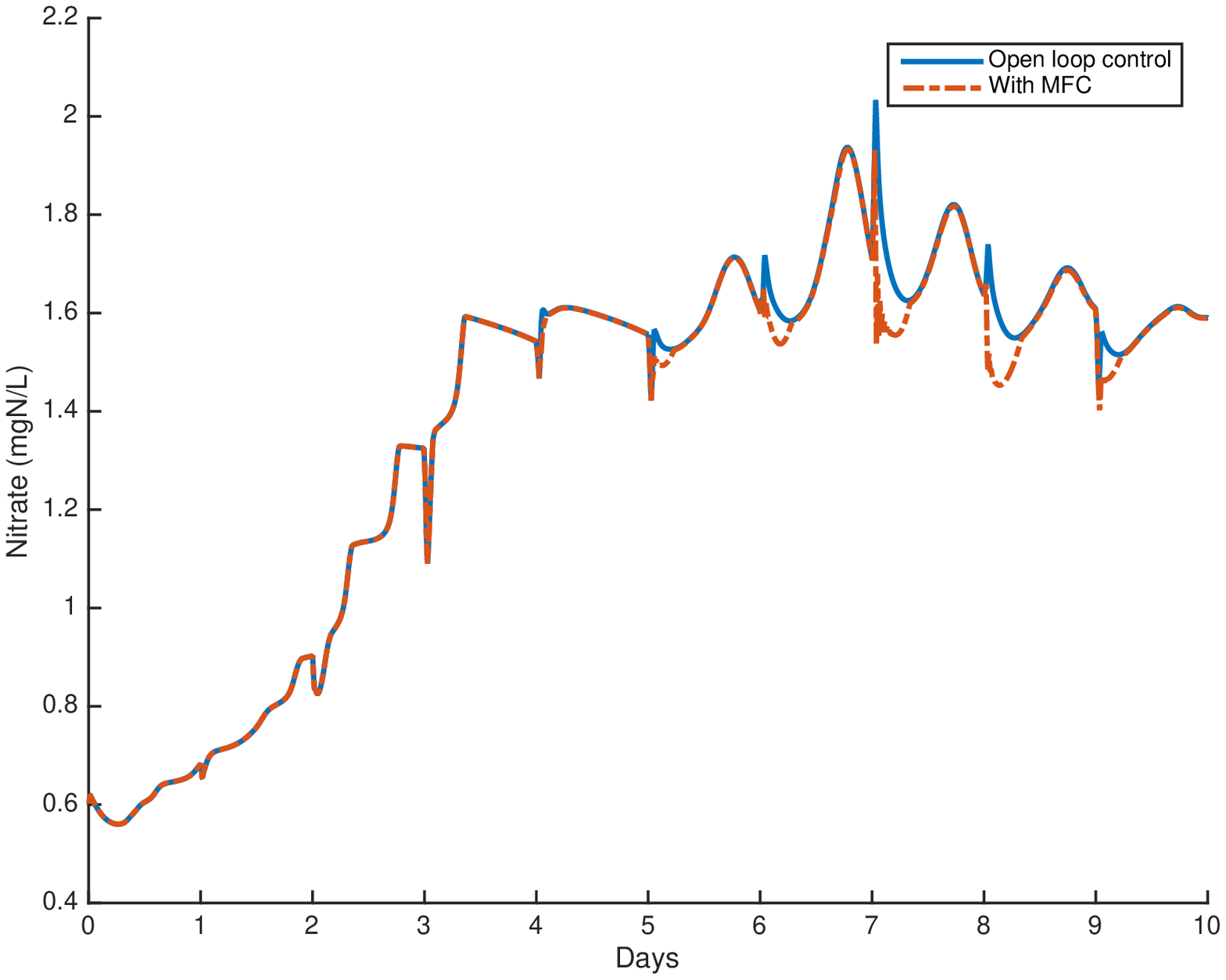}}}%
\subfigure[Nitrite $S_2$]{
\resizebox*{5.22cm}{!}{\includegraphics{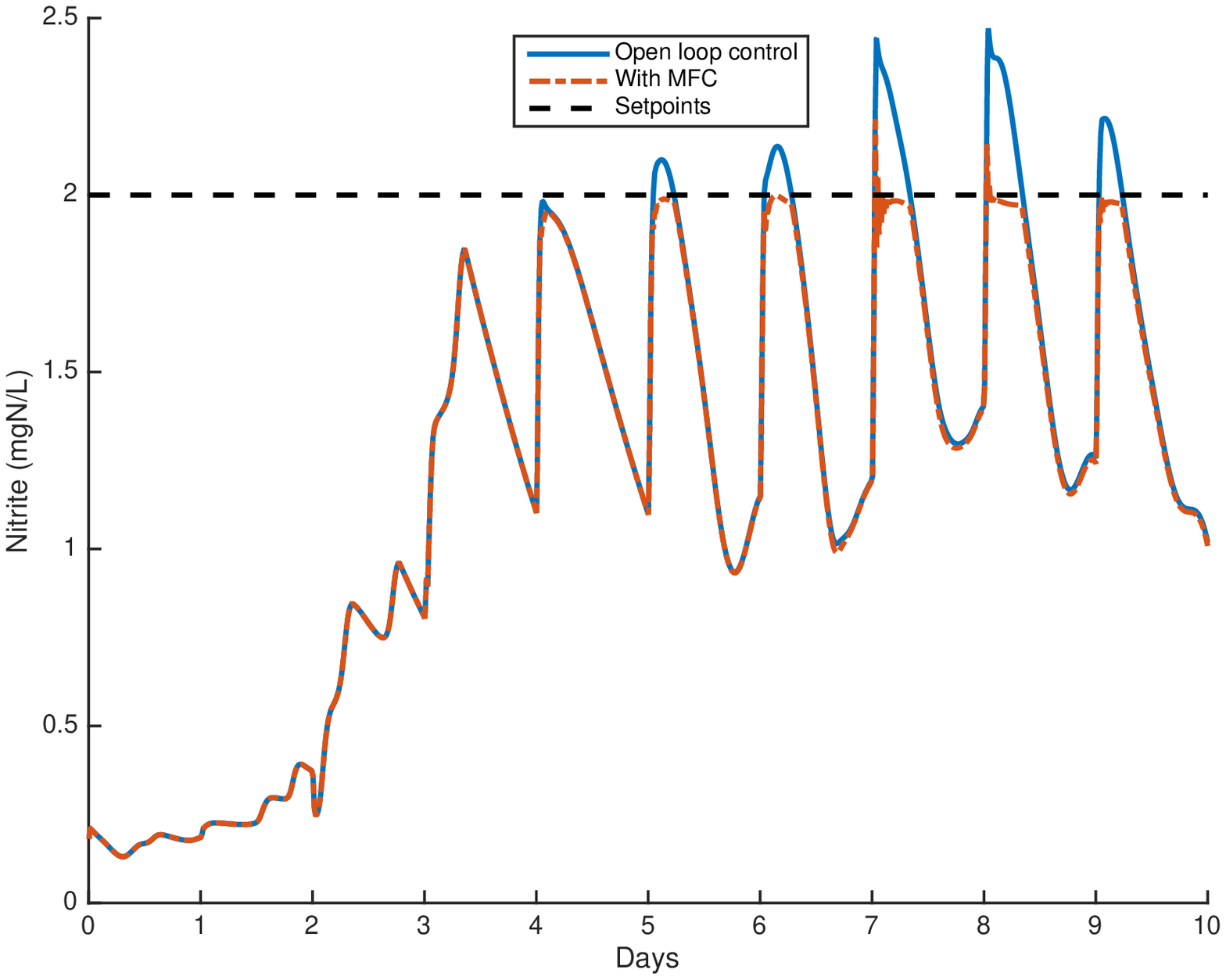}}}%
\subfigure[Methanol $S_{c,in}$]{
\resizebox*{5.22cm}{!}{\includegraphics{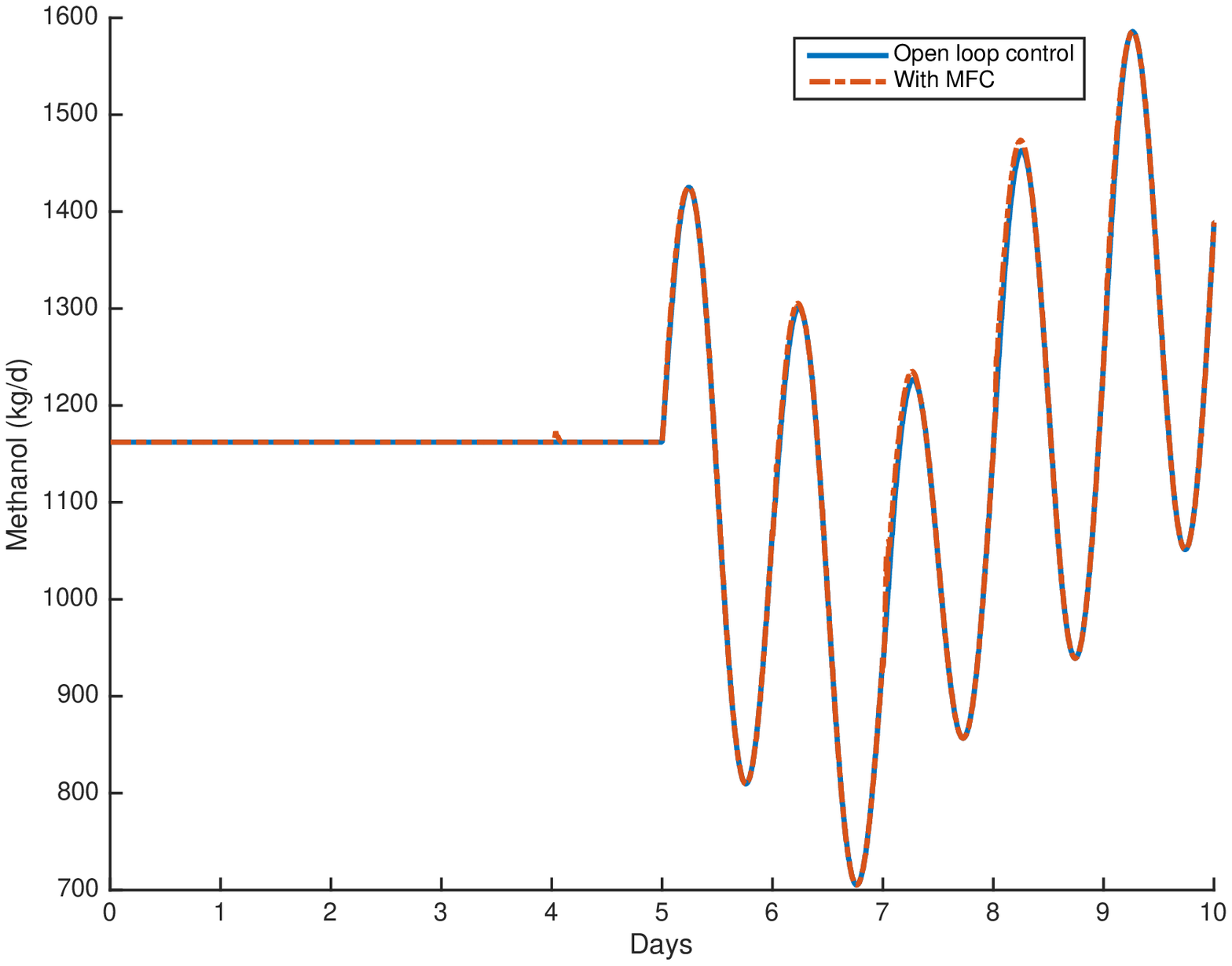}}}%
\caption{Setpoint : $S_{2,\text{target}}=2.0$ }%
\label{S20}
\end{center}
\end{figure*}

\begin{figure*}
\begin{center}
\subfigure[Nitrate $S_1$]{
\resizebox*{5.22cm}{!}{\includegraphics{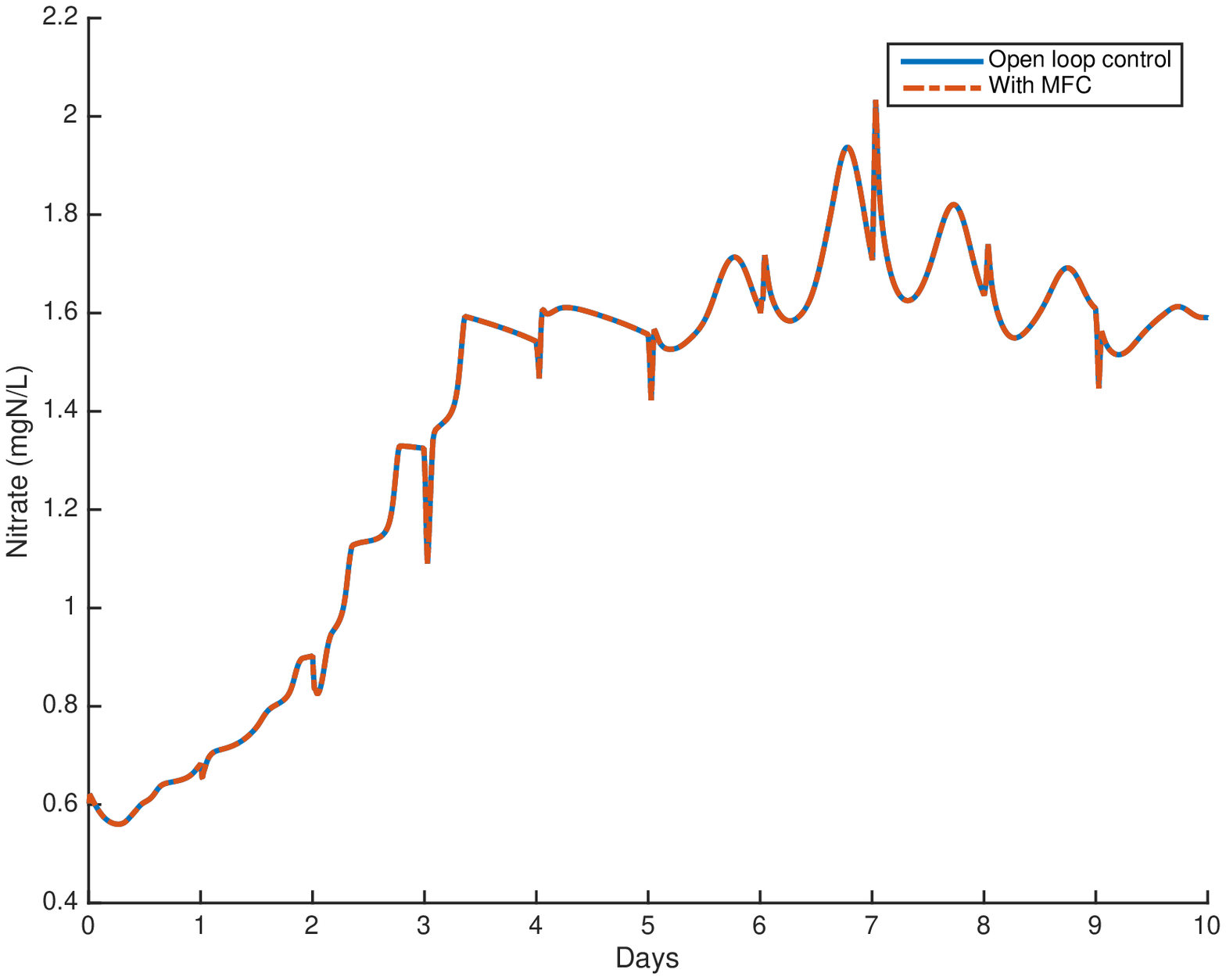}}}%
\subfigure[Nitrite $S_2$]{
\resizebox*{5.22cm}{!}{\includegraphics{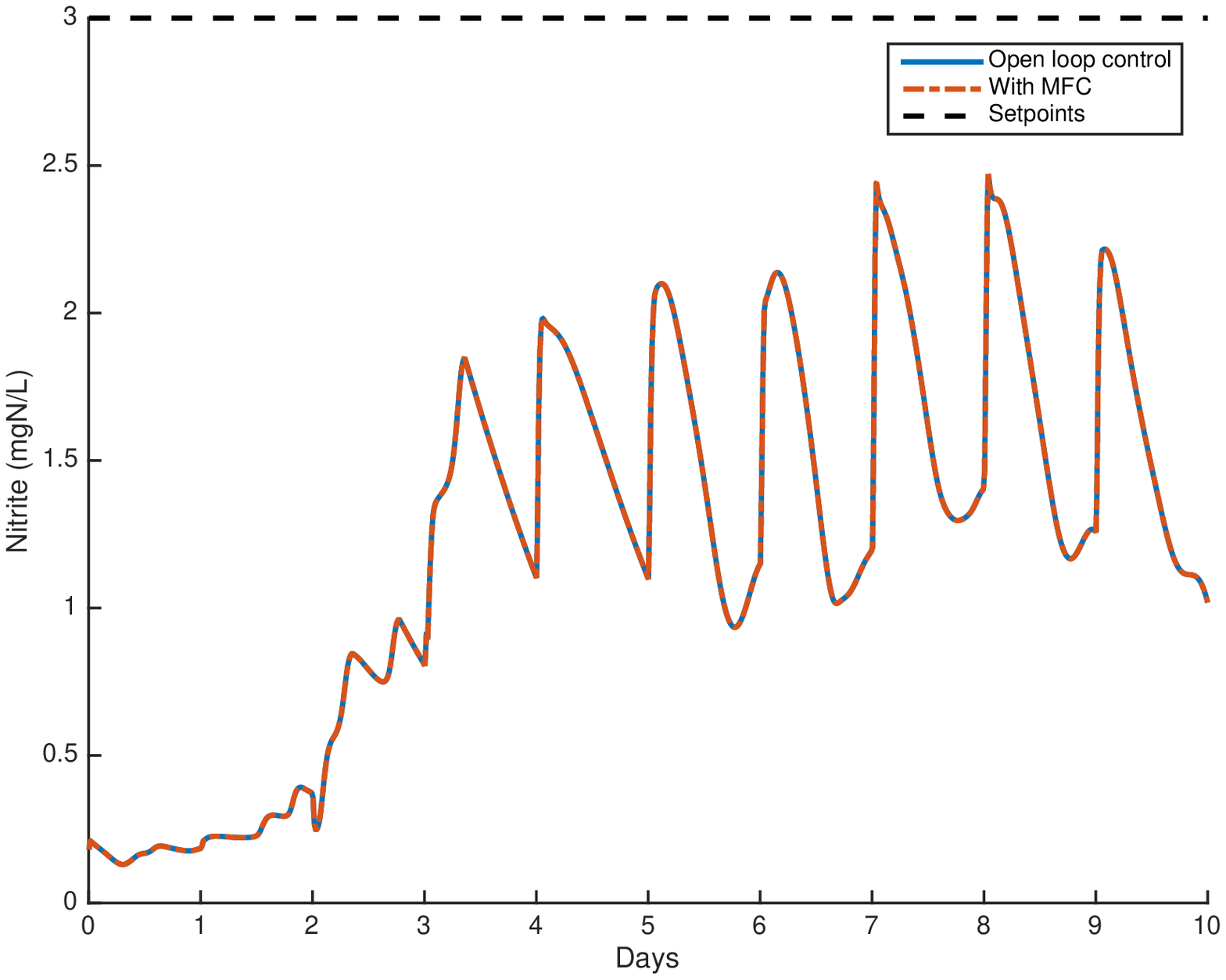}}}%
\subfigure[Methanol $S_{c,in}$]{
\resizebox*{5.22cm}{!}{\includegraphics{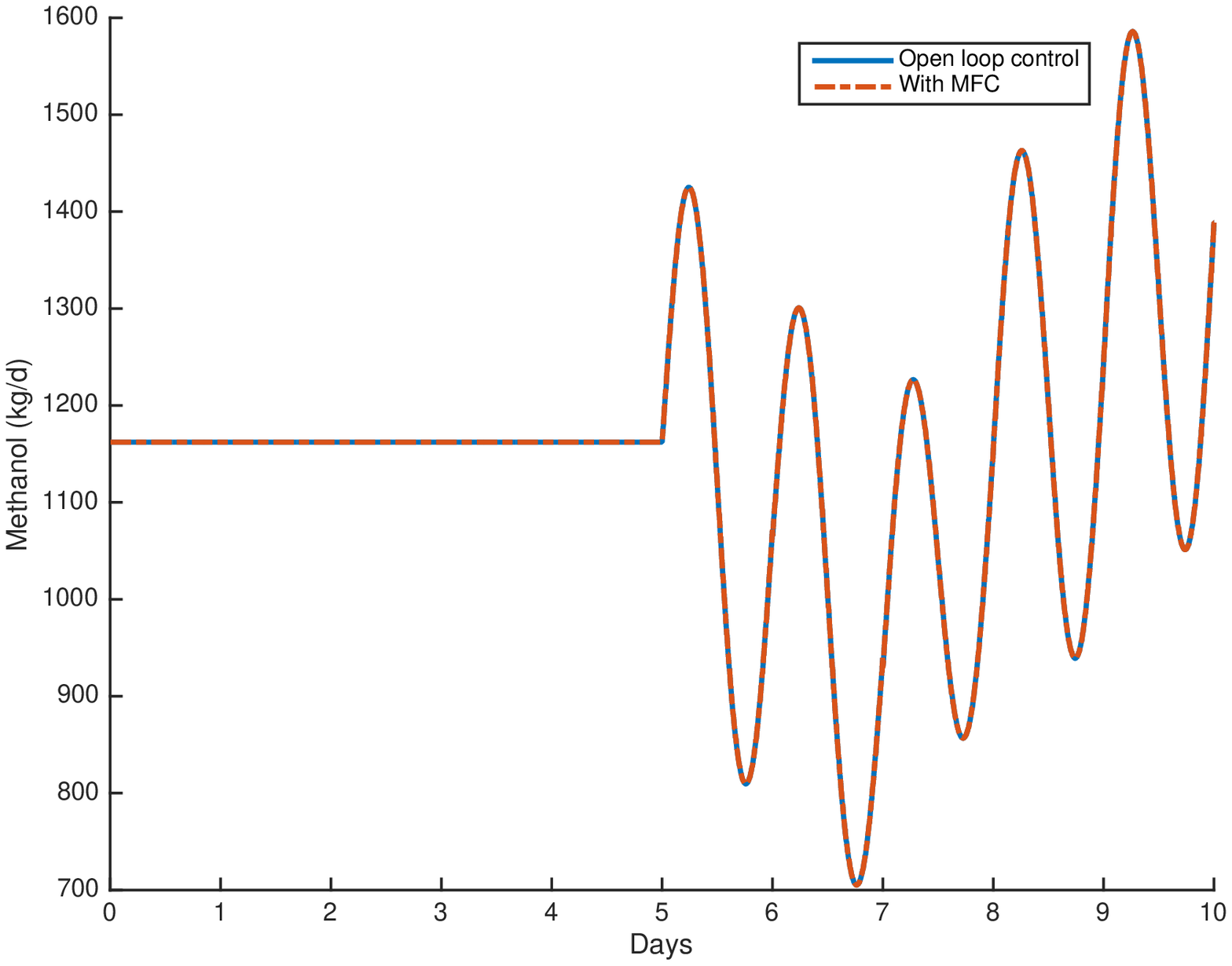}}}%
\caption{Setpoint : $S_{2,\text{target}}=3.0$ }%
\label{S30}
\end{center}
\end{figure*}

\section{Conclusion}\label{con}
This new setting for wastewater denitrification, which seems to be rather promising (see \cite{eau} for further details), will soon be tested in Paris. Although it has been shown in other applications that model-free control behaves quite well with respect to noise corruptions, it will provide us with more realistic data on noisy measurements, and other perturbations, for water treatment. The results will be reported elsewhere.

\begin{ack}
The authors are thankful to the research program \textit{MOCOP\'EE}\footnote{MOCOP\'EE is an acronym of  the French words `\underline{MO}d\'elisation \underline{C}ontrôle et \underline{O}ptimisation des \underline{P}roc\'ed\'es d'\underline{\'E}puration des \underline{E}aux.''} for its technical and financial support.
\end{ack}


\begin{thebibliography}{6}
\providecommand{\natexlab}[1]{#1}
\providecommand{\url}[1]{\texttt{#1}}
\providecommand{\urlprefix}{URL }
\expandafter\ifx\csname urlstyle\endcsname\relax
  \providecommand{\doi}[1]{doi:\discretionary{}{}{}#1}\else
  \providecommand{\doi}{doi:\discretionary{}{}{}\begingroup
  \urlstyle{rm}\Url}\fi
  
\bibitem[{{\AA}str\"om \textit{et~al.}(2006)}]{astrom}
{\AA}str\"om K.J., H\"agglund T. (2006). \emph{Advanced PID
Control}. Instrument Soc. Amer.



\bibitem[{{\AA}str\"om \textit{et~al.}(2008)}]{murray}
{\AA}str\"om K.J., Murray R.M. (2008). \emph{Feedback Systems: An Introduction for Scientists and Engineers}. Princeton University Press.


\bibitem[{Bara \textit{et~al.}(2016)}]{bara}
Bara O., Fliess M., Join C., Day J., Djouadi S.M. (2016). Model-free immune therapy: A control approach to acute inflammation. \emph{Europ. Contr. Conf.}, Aalborg. {\tt \small https://hal.archives-ouvertes.fr/hal-01341060/en/}

\bibitem[{Bastin \textit{et~al.}(1990)}]{bastin2013line}
Bastin G., Dochain D. (1990).
\newblock \emph{On-Line Estimation and Adaptive Control of Bioreactors}.
\newblock Elsevier.

\bibitem[{Bernier \textit{et~al.}(2014)Bernier, Rocher, and Lessard}]{Bernier2014}
Bernier J., Rocher V., Lessard P. (2014).
\newblock Modelling headloss and two-step denitrification in a full-scale
  wastewater post-denitrifying biofiltration plant.
\newblock \emph{J. Environ. Engin. Sci.}, 9,
  171--180.

\bibitem[{Bourrel \textit{et~al.}(2000)Bourrel, Dochain, Babary, and
  Queinnec}]{bourrel2000modelling}
Bourrel S., Dochain D., Babary J., Queinnec I. (2000).
\newblock Modelling, identification and control of a denitrifying biofilter.
\newblock \emph{J. Process Contr.}, 10, 73--91.

\bibitem[{Capodaglio \textit{et~al.}(2016)}]{capo}
Capodaglio A.G., Hlav\'{\i}nek P., Raboni M. (2016).
\newblock Advances in wastewater nitrogen removal by biological processes: state of the art review.
\newblock \emph{Ambiente \& \'Agua}, 11, 250--267.

\bibitem[{Cristea \textit{et~al.}(2008)}]{crist}
Cristea V.-M., Pop C., Agachi P.S. (2008). Model Predictive Control of the Waste Water Treatment Plant Based on the Benchmark Simulation Model No.1-BSM1. B. Braunschweig \& X. Joulia (Eds):
\emph{18th Europ. Symp. Comput. Aided Process Engin. - ESCAPE 18}, Elsevier.


\bibitem[Dochain \textit{et~al.}(2001)]{pde}
Dochain D., Vanrolleghem P.A. (2001). \emph{Dynamic Modelling and Estimation in Wastewater Treatment Processes}. IWA Publishing.

\bibitem[{Erd\'elyi(1962)}]{yosida}
Erd\'elyi A. (1962). \emph{Operational Calculus and Generalized Functions}. Holt Rinehart Winston.

\bibitem[Fliess(1989)]{diff}
Fliess M. (1989). Automatique et corps diff\'erentiels. \emph{Forum Math.}, 1, 227--238.

\bibitem[Fliess \textit{et~al.}(2013)]{ijc}
Fliess M., Join C. (2013). Model-free control.
\emph{Int. J. Contr.}, 86, 2228--2252.

\bibitem[Fliess \textit{et~al.}(2003)]{sira1}
Fliess M., Sira-Ram\'{\i}rez H. (2003).
\newblock An algebraic framework for linear identification.
\newblock \emph{ESAIM Contr. Optimiz. Calc. Variat.},
9, 151--168.

\bibitem[Fliess \textit{et~al.}(2008)]{sira2}
Fliess M., Sira-Ram\'{\i}rez H. (2008).
\newblock Closed-loop parametric identification for
  continuous-time linear systems via new algebraic techniques.
H. Garnier \& L. Wang (Eds): \emph{Identification of
  Continuous-time Models from Sampled Data}, Springer,
 pp. 362--391.
 
\bibitem[Fux \textit{et~al.}(2015)]{fux} 
Fux C., Kienle C., Joss A., Wittmer A., Frei R. (2015). Ausbau der ARA Basel mit 4. Reinigungsstufe. Pilotstudie: Elimination Mikroverunreinigungen und \"{o}kotoxikologische Wirkungen. 
\emph{Aqua \& Gas: Fachzeitschrift Gas Wasser Abwasser}, 97, 10--17.

\bibitem[Godement(1998)]{godement}
Godement R. (1998). \newblock \emph{Analyse mathématique II}, Springer  \newblock (English translation (2005): \emph{Analysis II}. Springer).
 
\bibitem[Grady Jr. \textit{et~al.}(2011)]{grad}
Grady Jr. C.P.L., Daigger G.T., Love N.G., Filipe C.D.M. (2011). \emph{Biological Wastewater Treatment} (3rd ed.). CRC Press.

\bibitem[Henze \textit{et~al.}(1987)]{henzeASM1} Henze M., Grady Jr. C.P.L., Gujer W., Marais G.V.R., Matsuo T.  (1987). \newblock A general-model for single-sludge waste-water treatment systems. \newblock \emph{Water Research}, 21, 505--515.

 \bibitem[Henze \textit{et~al.}(2008)]{henze} 
 Henze M., van Loosdrecht M.C.M., G. A. Ekama G.A., Brdjanovic D. (Eds) (2008). \emph{Biological Wastewater Treatment: Principles, Modeling, and Design}. IWA Publishing.

\bibitem[Hiatt \textit{et~al.}(2008)]{hiatt} Hiatt W.C., Grady C.P.L. (2008). \newblock An updated process model for carbon oxidation, nitrification, and denitrification. \newblock \emph{Water Environment Research}, 80, 2145--2156.

\bibitem[Horner \textit{et~al.}(1986)]{horner} Horner R.M.W, Jarvis R.J., Mackie R.I. (1986). \newblock Deep bed filtration - a new look at the basic equations. \newblock \emph{Water Research} 20, 215--220.

\bibitem[Ives (1970)]{ives} Ives K.J. (1970).\newblock Rapid filtration. \newblock\emph{Water Research} 4, 201--223.

\bibitem[Join \textit{et~al.}(2013)]{nice}
Join C., Chaxel F., Fliess M. (2013). ``Intelligent'' controllers on cheap and small programmable devices. \emph{2nd Int. Conf. Contr. Fault-Tolerant Syst. (SysTol'13)}, Nice. 
{\tt \small https://hal.archives-ouvertes.fr/hal-00845795/en/}


\bibitem[Join \textit{et~al.}(2010)]{edf}
Join C., Robert G., Fliess M. (2010). Vers une commande sans
mod\`{e}le pour am\'{e}nagements hydro\'{e}lectriques en cascade,
\emph{6$^e$ Conf. Internat. Francoph. Automat.}, Nancy. \newline
{\tt \small http://hal.archives-ouvertes.fr/inria-00460912/en/}

\bibitem[{Lafont \textit{et~al.}(2015)}]{toulon}
Lafont F., Balmat J.-F., Pessel N., Fliess M. (2015).
A model-free control strategy for an experimental greenhouse with an application to fault accommodation. \emph{Comput. Electron. Agricult.}, 110, 139--149.


\bibitem[{Marsili-Libelli  \textit{et~al.}(2002)}]{predict}
Marsili-Libelli S., Giunti L. (2002). Fuzzy predictive control for nitrogen removal in biological wastewater treatment. \emph{Water Sci. Techno.}, 45, 37--44.


\bibitem[{Olsson  \textit{et~al.}(1999)}]{ols}
Olsson G., Newell B. (1999). \emph{Wastewater treatment systems: modelling, diagnosis and control}. IWA Publishing.

\bibitem[{Raimonet \textit{et~al.}(2015)Raimonet, Vilmin, Flipo, Rocher, and
  Laverman}]{Raimonet2015373}
Raimonet M., Vilmin L., Flipo N., Rocher V., Laverman A.M. (2015).
\newblock Modelling the fate of nitrite in an urbanized river using
  experimentally obtained nitrifier growth parameters.
\newblock \emph{Water Res.}, 73, 373 -- 387.

\bibitem[{Rocher \textit{et~al.}(2017)}]{eau}
Rocher V., Join C., Mottelet S., Bernier J., Gu\'{e}rin S., Azimi S.,  Lessard P., Pauss A., Fliess M. (2017).
\newblock{La production de nitrites lors de la dénitrification des eaux usées par biofiltration - Stratégie de contr\^{o}le et de réduction des concentrations résiduelles}.
\newblock \emph{Rev. Sci. Eau}, submitted.

\bibitem[{Rocher \textit{et~al.}(2015)Rocher, Laverman, Gasperi, Azimi, Gu{\'e}rin,
  Mottelet, Villi{\`e}res, and Pauss}]{Rocher2015}
Rocher V., Laverman A.M., Gasperi J., Azimi S., Gu{\'e}rin S., Mottelet
  S., Villi{\`e}res T., Pauss A. (2015).
\newblock Nitrite accumulation during denitrification depends on the carbon
  quality and quantity in wastewater treatment with biofilters.
\newblock \emph{Environ. Sci. Pollut. Res.}, 22,
  10179--10188.

\bibitem[{Samie \textit{et~al.}(2011)}]{Samie2011}
Samie G., Bernier J., Rocher V., Lessard P. (2011).
\newblock Modeling nitrogen removal for a denitrification biofilter.
\newblock \emph{Biopro. Biosyst. Engin.}, 34, 747--755.

\bibitem[{Sira-Ram\'{\i}rez \textit{et~al.}(2014)}]{sira}
Sira-Ram\'{\i}rez H., Garc\'{\i}a-Rodr\'{\i}guez C., Cort\`{e}s-Romero  J., Luviano-Ju\'{a}rez A. (2013). \emph{Algebraic Identification and Estimation Methods in Feedback Control Systems}. Wiley.

\bibitem[Spengel \textit{et~al.}(1992)]{spengel} Spengel D., Dzombak D. (1992). \newblock Biokinetic modeling and scale-up considerations for rotating biological contactors. \newblock \emph{Water Environ. Res.} 64, 223-235. 

\bibitem[{Tebbani \textit{et~al.}(2016)}]{med16}
Tebbani S., Titica M., Join C., Fliess M., Dumur D. (2016).
Model-based versus model-free control designs
for improving microalgae growth in a closed photobioreactor: Some preliminary comparisons. \emph{24th Medit. Conf. Contr. Automat.}, Athens. \newline {\tt \small https://hal.archives-ouvertes.fr/hal-01312251/en/}

\bibitem[{Torres Z\'{u}ñiga \textit{et~al.}(2012)}]{tor}
Torres Z\'{u}$\tilde{\text n}$iga I., Queinnec I., Vande Wouwer A. (2012). 
\newblock Observer-based output feedback linearizing control strategy for a nitrification-denitrification biofilter.
\newblock \emph{Chem. Engin. J.}, 191, 243-255.


\bibitem[{Wahaba \textit{et~al.}(2009)}]{wah}
Wahaba N.A., Katebia R., Balderud J. (2009).
\newblock  Multivariable PID control design for activated sludge process with nitrification and denitrification.
\newblock \emph{Biochem. Engin. J.}, 45, 239--248.

\bibitem[{Water Environment Federation(2013)}]{wef}
Water Environment Federation (2013). \emph{Wastewater Treatment Process Modeling}. McGraw-Hill.


\end{thebibliography}

\end{document}